\documentclass[11pt]{myArticle}       

\usepackage{amsmath,amssymb,amsfonts} 
\usepackage{graphicx}
\usepackage{epsfig}
\usepackage{color}                    
\usepackage{hyperref}                 
\usepackage{myFancyhdr}               
\usepackage{makeidx}                  
\usepackage{helvet}                   
\usepackage{setspace}                 
\usepackage{fancybox}                 
\usepackage{float}                    
\usepackage{xspace}                   
\usepackage{wasysym}                  
\usepackage[numbers,sort&compress]{natbib}
\usepackage{hypernat}
\usepackage{wrapfig}

\definecolor{mygray}{rgb}{0.3,0.32,0.35}
\definecolor{darkblue1}{rgb}{0,0,.2}
\definecolor{darkblue}{rgb}{0,0,.3}
\definecolor{darkred}{rgb}{0.5,0,0}
\pagecolor{white} 
\color{black}     
\hypersetup{breaklinks=true, 
            colorlinks=true, 
            linkcolor=darkblue1, 
            menucolor=darkblue1, 
            urlcolor=darkblue1,
            citecolor=darkblue1,
            pdftitle={The SM EW fit Revisited},
            pdfauthor={Gfitter Group},
            pdfsubject={The SM EW fit Revisited},
            pdfkeywords={},
            pdfproducer={Gfitter Group}
}
\bibstyle{plain}
\renewcommand{\headrulewidth}{0.5pt}    
\renewcommand{\footrulewidth}{0.5pt}    

\newcommand\allFontSize{\small}

\newenvironment{myquote}
               {\list{}{\leftmargin0cm}%
                \item\relax}
               {\endlist}

\usepackage[bf]{caption}

\newcommand\detailsSize{\allFontSize}
\newenvironment{details}%
{\begin{myquote}\vspace{-0.2cm}\detailsSize}{\end{myquote}\vspace{-0.2cm}}

\newfloat{codeexample}{H}{loc}
\floatname{codeexample}{\sf\allFontSize Code Example}

\newlength{\gfitterboxwidth}
\definecolor{DarkGray}{rgb}{0.4,0.42,0.45}
\definecolor{LightGray}{rgb}{0.97,0.98,0.98}

{\VerbatimEnvironment\normalsize
\setlength{\gfitterboxwidth}{\textwidth}\addtolength{\gfitterboxwidth}{-2.3\fboxsep}%
\begin{Sbox}\begin{minipage}[t]{\gfitterboxwidth}\begin{Verbatim}}%
{\end{Verbatim}\end{minipage}\end{Sbox}\vspace{1ex} \fcolorbox{DarkGray}{LightGray}{\TheSbox}}

\newfloat{option}{H}{loc}
\floatname{option}{\sf\allFontSize Option Table}

\usepackage{tabularx}
{\setlength{\gfitterboxwidth}{\textwidth}\addtolength{\gfitterboxwidth}{-2.3\fboxsep}%
\begin{Sbox}\begin{tabular*}{\gfitterboxwidth}{>{\rule[-0.5ex]{.0ex}{3.0ex}\tt\small}p{3cm}>{\tt\small}p{4.5cm}>{\small}p{5.7cm}}
&&\\[-0.7cm]
\rm\small Option  & \rm\small Values    & \rm\small Description\\[0.1cm]\hline
&&\\[-0.45cm]}%
{\\[-0.1cm]\end{tabular*}\end{Sbox}\vspace{1ex} \fbox{\TheSbox}}

{\setlength{\gfitterboxwidth}{\textwidth}\addtolength{\gfitterboxwidth}{-2.3\fboxsep}%
\begin{Sbox}\begin{tabular*}{\gfitterboxwidth}{>{\rule[-0.5ex]{.0ex}{3.0ex}\tt\small}p{3cm}>{\tt\small}p{3.5cm}>{\small}p{6.7cm}}
&&\\[-0.7cm]
\rm\small Option  & \rm\small Values    & \rm\small Description\\[0.1cm]\hline
&&\\[-0.45cm]}%
{\\[-0.1cm]\end{tabular*}\end{Sbox}\vspace{1ex} \fbox{\TheSbox}}

{\setlength{\gfitterboxwidth}{\textwidth}\addtolength{\gfitterboxwidth}{-2.3\fboxsep}%
\begin{Sbox}\begin{tabular*}{\gfitterboxwidth}{>{\rule[-0.5ex]{.0ex}{3.5ex}\tt\small}p{3.0cm}>{\tt\small}p{3.0cm}>{\small}p{7.2cm}}
&&\\[-0.7cm]
\rm\small Option  & \rm\small Values    & \rm\small Description\\[0.1cm]\hline
&&\\[-0.45cm]}%
{\\[-0.1cm]\end{tabular*}\end{Sbox}\vspace{1ex} \fbox{\TheSbox}}

{\setlength{\gfitterboxwidth}{\textwidth}\addtolength{\gfitterboxwidth}{-2.3\fboxsep}%
\begin{Sbox}\begin{tabular*}{\gfitterboxwidth}{>{\rule[-0.5ex]{.0ex}{3.0ex}\tt\small}p{2.5cm}>{\tt\small}p{3.5cm}>{\small}p{7.2cm}}
&&\\[-0.7cm]
\rm\small Option  & \rm\small Values    & \rm\small Description\\[0.1cm]\hline
&&\\[-0.45cm]}%
{\\[-0.1cm]\end{tabular*}\end{Sbox}\vspace{1ex} \fbox{\TheSbox}}

{\setlength{\gfitterboxwidth}{\textwidth}\addtolength{\gfitterboxwidth}{-2\fboxsep}%
\begin{Sbox}\begin{tabular*}{\gfitterboxwidth}{>{\rule[-0.5ex]{.0ex}{3.0ex}\tt\small}p{4.5cm}>{\tt\small}p{3.2cm}>{\small}p{5.5cm}}
&&\\[-0.7cm]
\rm\small Option  & \rm\small Values    & \rm\small Description\\[0.1cm]\hline
&&\\[-0.45cm]}%
{\\[-0.1cm]\end{tabular*}\end{Sbox}\vspace{1ex} \fbox{\TheSbox}}

{\setlength{\gfitterboxwidth}{\textwidth}\addtolength{\gfitterboxwidth}{-2\fboxsep}%
\begin{Sbox}\begin{tabular*}{\gfitterboxwidth}{>{\rule[-0.5ex]{.0ex}{3.0ex}\tt\small}p{4.0cm}>{\tt\small}p{3.0cm}>{\small}p{6.2cm}}
&&\\[-0.7cm]
\rm\small Option  & \rm\small Values    & \rm\small Description\\[0.1cm]\hline
&&\\[-0.45cm]}%
{\\[-0.1cm]\end{tabular*}\end{Sbox}\vspace{1ex} \fbox{\TheSbox}}

\newfloat{programs}{H}{loc}
\floatname{programs}{\sf Table}

\usepackage{tabularx}
{\setlength{\gfitterboxwidth}{\textwidth}\addtolength{\gfitterboxwidth}{-2\fboxsep}%
\begin{Sbox}\begin{tabular*}{\gfitterboxwidth}{>{\rule[-0.5ex]{.0ex}{3.0ex}\tt\small}p{4.1cm}>{\small}p{9.5cm}}
&\\[-0.7cm]
\rm\small Macro & \rm\small Description\\[0.1cm]\hline
&\\[-0.45cm]}%
{\\[-0.1cm]\end{tabular*}\end{Sbox}\vspace{1ex} \fbox{\TheSbox}}

\mathchardef\Upsilon="7107
\def\Y#1S{\ensuremath{\Upsilon{(#1S)}}\xspace}

\newcommand{\as}{\ensuremath{\alpha_{\scriptscriptstyle S}}\xspace}

\newcommand{\asZ}{\ensuremath{\as(M_Z^2)}\xspace}

\renewcommand\l{\ell}

\newcommand{\Kbar    }{\kern 0.2em\overline{\kern -0.2em K}{}\xspace}

\newcommand{\Kz      }{\ensuremath{K^0}\xspace}
\newcommand{\Kzb     }{\ensuremath{\Kbar^0}\xspace}
\newcommand{\KzKzb   }{\ensuremath{\Kz \kern -0.16em \Kzb}\xspace}
\newcommand{\Kp      }{\ensuremath{K^+}\xspace}
\newcommand{\Km      }{\ensuremath{K^-}\xspace}

\newcommand{\KpKm    }{\ensuremath{\Kp \kern -0.16em \Km}\xspace}

\newcommand\Dbar    {\kern 0.18em\overline{\kern -0.18em D}{}\xspace}

\newcommand\Bbar    {\kern 0.18em\overline{\kern -0.18em B}{}\xspace}

\newcommand\Bz      {\ensuremath{B^0}\xspace}

\newcommand\Bzb     {\ensuremath{\Bbar^0}\xspace}
\newcommand\Bu      {\ensuremath{B^+}\xspace}
\newcommand\Bub     {\ensuremath{B^-}\xspace}

\newcommand\BpBm    {\ensuremath{\Bu {\kern -0.16em \Bub}}\xspace}
\newcommand\Bs      {\ensuremath{B^0_{s}}\xspace}
\newcommand\Bsb     {\ensuremath{\Bbar^0_{s}}\xspace}

\newcommand\BzBzb   {\ensuremath{\Bz {\kern -0.16em \Bzb}}\xspace}
\newcommand\BszBszb {\ensuremath{\Bs {\kern -0.16em \Bsb}}\xspace}

\newcommand\deltatheo{\ensuremath{\delta_{\rm th}}\xspace}

\newcommand\Rfit{{\em R}fit\xspace}

\newcommand{\ft}{\footnotesize}

\newcommand{\tev}{\ensuremath{\mathrm{Te\kern -0.1em V}}\xspace}
\newcommand{\gev}{\ensuremath{\mathrm{Ge\kern -0.1em V}}\xspace}
\newcommand{\mev}{\ensuremath{\mathrm{Me\kern -0.1em V}}\xspace}
\newcommand{\kev}{\ensuremath{\mathrm{ke\kern -0.1em V}}\xspace}
\newcommand{\ev}{\ensuremath{\mathrm{e\kern -0.1em V}}\xspace}
\newcommand{\gevc}{\ensuremath{{\mathrm{Ge\kern -0.1em V\!/}c}}\xspace}
\newcommand{\mevc}{\ensuremath{{\mathrm{Me\kern -0.1em V\!/}c}}\xspace}
\newcommand{\gevcc}{\ensuremath{{\mathrm{Ge\kern -0.1em V\!/}c^2}}\xspace}
\newcommand{\mevcc}{\ensuremath{{\mathrm{Me\kern -0.1em V\!/}c^2}}\xspace}

\newcommand{\bei}{\begin{itemize}}
\newcommand{\eei}{\end{itemize}}
\newcommand{\beq}{\begin{equation}}
\newcommand{\eeq}{\end{equation}}
\newcommand{\beqn}{\begin{eqnarray}}
\newcommand{\eeqn}{\end{eqnarray}}
\newcommand{\beqns}{\begin{eqnarray*}}
\newcommand{\eeqns}{\end{eqnarray*}}
\newcommand{\bitm}{\begin{itemize}}
\newcommand{\eitm}{\end{itemize}}

\newcommand{\dahadZf}{\ensuremath{\Delta\alpha_{\rm had}^{(5)}(M_Z^2)}\xspace}
\newcommand{\dalphaHadMZ}{\ensuremath{\Delta\alpha_{\rm had}^{(5)}(M_Z^2)}\xspace}

\newcommand\ie{{i.e.}\xspace}

\newcommand\cf{{cf.}\xspace}

\newcommand\rs{\raisebox{1.5ex}[-1.5ex]}

\newcommand{\STU}{$S,\,T,\,U$\xspace}
\def\@citex[#1]#2{\if@filesw\immediate\write\@auxout{\string\citation{#2}}\fi
  \@tempcnta\z@\@tempcntb\m@ne\def\@citea{}\@cite{\@for\@citeb:=#2\do
    {\@ifundefined
       {b@\@citeb}{\@citeo\@tempcntb\m@ne\@citea
        \def\@citea{,\penalty\@m\ }{\bf ?}\@warning
       {Citation `\@citeb' on page \thepage \space undefined}}%
    {\setbox\z@\hbox{\global\@tempcntc0\csname b@\@citeb\endcsname\relax}%
     \ifnum\@tempcntc=\z@ \@citeo\@tempcntb\m@ne
       \@citea\def\@citea{,\penalty\@m}
       \hbox{\csname b@\@citeb\endcsname}%
     \else
      \advance\@tempcntb\@ne
      \ifnum\@tempcntb=\@tempcntc
      \else\advance\@tempcntb\m@ne\@citeo
      \@tempcnta\@tempcntc\@tempcntb\@tempcntc\fi\fi}}\@citeo}{#1}}

\def\@citeo{\ifnum\@tempcnta>\@tempcntb\else\@citea
  \def\@citea{,\penalty\@m}%
  \ifnum\@tempcnta=\@tempcntb\the\@tempcnta\else
   {\advance\@tempcnta\@ne\ifnum\@tempcnta=\@tempcntb \else
\def\@citea{--}\fi
    \advance\@tempcnta\m@ne\the\@tempcnta\@citea\the\@tempcntb}\fi\fi}

\xspace

\newcommand\mini{{\rm min}}

\newcommand\ChiMin{\ensuremath{\chi^2_{\mini}}\xspace}
\newcommand\DeltaChi{\ensuremath{\Delta\chi^2}\xspace}

\newcommand{\seffsf}[1]{\sin\!^2\theta^{#1}_{{\rm eff}}}
\newcommand{\sinleff}{\seffsf{\ell}}
\newcommand{\mc}{\ensuremath{\overline{m}_c}\xspace}
\newcommand{\mb}{\ensuremath{\overline{m}_b}\xspace}

\newcommand{\WMass}     	{\ensuremath{80.385\pm0.015\:\gev}}
\newcommand{\WWidth}     	{\ensuremath{2.085\pm0.042\:\gev}}
\newcommand{\TopMass}     	{\ensuremath{173.18\pm0.94\:\gev}}

\newcommand{\ChiMinVal}     	{\ensuremath{21.8}\xspace}
\newcommand{\NDFVal}     	{\ensuremath{14}\xspace}
\newcommand{\ProbVal}     	{\ensuremath{0.07}\xspace}

\newcommand{\WMassInd}     	{\ensuremath{80.359\pm0.011\:\gev}}

\newcommand{\TopMassInd}        {\ensuremath{175.8^{\:+2.7}_{\:-2.4}\:\gev}}

\newcommand{\SinSqInd}     	{\ensuremath{0.23150\pm0.00010}\xspace}

\newcommand{\SParam}     	{\ensuremath{0.03\pm 0.10}\xspace}
\newcommand{\TParam}     	{\ensuremath{0.05\pm 0.12}\xspace}
\newcommand{\UParam}     	{\ensuremath{0.03\pm 0.10}\xspace}

\newcommand{\STParamCor}	{\ensuremath{+0.89}\xspace}
\newcommand{\SUParamCor}	{\ensuremath{-0.54}\xspace}
\newcommand{\TUParamCor}	{\ensuremath{-0.83}\xspace}

\newcommand{\SParamNU}     	{\ensuremath{0.05\pm 0.09}\xspace}
\newcommand{\TParamNU}     	{\ensuremath{0.08\pm 0.07}\xspace}
\newcommand{\STParamCorNo}     	{\ensuremath{+0.91}\xspace}

\makeindex
\begin{document}
%
%
\pagenumbering{arabic}
{\small
\color{mygray}
\begin{flushright}
{\sf\em CERN-OPEN-2012-xxx} \\
{\sf\em DESY-12-154} \\
{\sf\em \today} \\
\def\UrlFont{\sf\em}
\url{http://cern.ch/gfitter} 
\end{flushright}
}
\def\UrlFont{\rm}

\vspace{1.3cm}

{\sf\LARGE\bfseries
The Electroweak Fit of the Standard Model \\[0.25cm]
after the Discovery of a New Boson at the LHC
}

\vspace{1.0cm}

{\Large \em 
  The Gfitter Group \\[0.2cm]
}
{\Large
  M.~Baak$^{a}$, M.~Goebel$^{b}$, J.~Haller$^{c}$, A.~Hoecker$^{a}$, D.~Kennedy$^{b}$,  \\
  R.~Kogler$^{c}$, K.~M\"onig$^{b}$, M.~Schott$^{a}$, J.~Stelzer$^{d}$
}

\vspace{0.5cm}

{\normalsize
  $^{a}$CERN, Geneva, Switzerland \\
  $^{b}$DESY, Hamburg and Zeuthen, Germany \\ 
  $^{c}$Institut f\"ur Experimentalphysik, Universit\"at Hamburg, Germany\\
  $^{d}$Department of Physics and Astronomy, Michigan State University, East Lansing, USA

\vspace{1.0cm}

\begin{details}
{\sf\bfseries Abstract} --- In view of the discovery of a new boson
by the ATLAS and CMS Collaborations at the LHC, we present an update
of the global Standard Model (SM) fit to electroweak precision data.
Assuming the new particle to be the SM Higgs boson, all fundamental 
parameters of the SM are known allowing, for the first time, to 
overconstrain the SM at the electroweak scale and assert its validity. 
Including the effects of radiative corrections and the experimental and 
theoretical uncertainties, the global fit exhibits a $p$-value of 0.07. 
The mass measurements by ATLAS and CMS agree within 1.3$\sigma$ with the
indirect determination $M_H=94^{\,+25}_{\,-22}$\,GeV. 
Within the SM the $W$ boson mass and the effective weak mixing angle 
can be accurately predicted to be 
$M_W=\WMassInd$ and $\sinleff=\SinSqInd$ from the global fit. These results are compatible 
with, and exceed in precision, the direct measurements. For the indirect 
determination of the top quark mass we find $m_t=\TopMassInd$, in 
agreement with the kinematic and cross-section based measurements.
\end{details}

\thispagestyle{empty}

\newpage
%
%
\section{Introduction}
\label{sec:intro}

The discovery by the ATLAS~\cite{:2012gk} and CMS~\cite{:2012gu} experiments at 
the LHC of a new particle with mass $\sim$126\:\gev and with properties compatible 
with those of the Standard Model (SM) Higgs boson 
concludes decades of intense experimental and theoretical work 
to uncover the mechanism of electroweak symmetry breaking and mass generation. 
If forthcoming data confirm that the new particle is the SM Higgs boson, this discovery 
exhibits another -- possibly the greatest ever -- triumph of the SM, as not 
only the SM predicts the Higgs couplings to the SM fermions and bosons, but it also 
constrains the Higgs boson to be light compared to its unitarity bound of roughly a TeV. 
This indirect information on the Higgs mass was extracted from Higgs loops  
affecting the values of $Z$ boson asymmetry observables and the $W$ mass. Global 
fits to precisely measured electroweak data derived 
95\% confidence level (CL) upper limits on the Higgs mass of around
160\:\gev~\cite{ALEPH:2010aa,LEPEWWG,ErlerNakamura:2010zzi,Baak:2011ze}.

In this letter we interpret the new particle as the SM Higgs boson 
and present the consequences on the global electroweak fit. 
A detailed description of the experimental data, the theoretical
calculations, and the statistical framework used in the analysis is
provided in past publications~\cite{Flacher:2008zq,Baak:2011ze}. Here,
we only briefly recall the most relevant aspects of the analysis and
highlight recent changes. The main goal of this letter is to quantify
the compatibility of the mass of the discovered boson with the electroweak
precision data and its impact on the indirect determination of the $W$ boson
mass, the effective weak mixing angle, and the top quark mass. The
implications of the discovery on the SM with three and four fermion
generations were also studied in \cite{Eberhardt:2012np}.

\section{Experimental data and theoretical predictions}
\label{sec:input}

The experimental inputs used in the fit include the electroweak
precision data measured at the $Z$ pole and their correlations~\cite{:2005ema}, 
the latest world average values for the mass of the $W$ boson, $M_W=
\WMass$~\cite{TevatronElectroweakWorkingGroup:2012gb}, and its width,
$\Gamma_W= \WWidth$~\cite{TEW:2010aj}, the latest average of the
direct top mass measurements from the Tevatron experiments, $m_t=
\TopMass$~\cite{Aaltonen:2012ra},\footnote
{
  The theoretical uncertainties arising from nonperturbative colour-reconnection 
  effects in the fragmentation process~\cite{Skands:2007zg,Wicke:2008iz}, and 
  from ambiguities in the top-mass definition~\cite{Hoang:2008yj,Hoang:2008xm}, 
  affect the (kinematic) top mass measurement. Their quantitative estimate is 
  difficult and may reach roughly 0.5\:\gev each, where the systematic error 
  due to shower effects could be larger~\cite{Skands:2007zg}. To estimate 
  the effect of a theoretical uncertainty of 0.5\:\gev, inserted in the fit as
  a uniform likelihood function according to the \Rfit scheme~\cite{Hocker:2001xe,Charles:2004jd},
  we have repeated the indirect determination of some of the most relevant 
  observables. We find in particular
  $M_H=90^{\,+34}_{\,-21}\:\gev$, $M_W=80.359 \pm 0.013\:\gev$, and 
  $\sinleff=0.23148 \pm 0.00010$, which, compared to the standard results given in 
  Eq.~(\ref{eq:mh}), (\ref{eq:mw}) and (\ref{eq:sin2t}), exhibit only a small
  deterioration in precision. 
} 
and the hadronic contribution 
to the running of the electromagnetic coupling strength, 
$\dahadZf=(2757\pm10)\cdot10^{-5}$~\cite{Davier:2010nc}. 
For the Higgs boson mass we use the measurements 
from ATLAS, $M_H=126.0\pm0.4\pm0.4\:\gev$~\cite{:2012gk},
and CMS, $M_H=125.3\pm0.4\pm0.5\:\gev$~\cite{:2012gu}, where the first 
uncertainties are statistical and the second systematic. 
A detailed list of all the observables, their values and uncertainties as used in the
fit, is given in the first two columns of Table~\ref{tab:results}.

A proper average of the Higgs mass measurements requires a detailed experimental study of 
the systematic correlations. Owing to the weak (logarithmic) dependence
of the electroweak fit on the Higgs mass, we find that the fit results are 
insensitive to the difference between a straight (uncorrelated) weighted 
average of the ATLAS and CMS  measurements ($M_H=125.7 \pm 0.4$\,GeV), 
and their weighted average obtained assuming the systematic uncertainties 
to be fully correlated  ($M_H=125.7 \pm 0.5$\,GeV).
In this paper the former combination is used.\footnote
{
  The main source of systematic uncertainty in the ATLAS
  and CMS mass measurements stems from the
  energy and momentum calibrations, which should be uncorrelated 
  between the experiments. 
  }

For the theoretical predictions, we use the calculations detailed
in~\cite{Baak:2011ze} and references therein. They feature among
others the complete $\mathcal{O}(\as^4)$ calculation of the QCD Adler
function~\cite{Baikov:2008jh, Baikov:2012er} and the full two-loop and leading
beyond-two-loop prediction of the $W$ mass and the effective weak
mixing angle~\cite{Awramik:2003rn,Awramik:2006uz,Awramik:2004ge}. 
Two modifications apply here:
first, an improved prediction of $R^0_b$ is invoked that includes the
calculation of the complete fermionic electroweak two-loop (NNLO) corrections based on
numerical Mellin-Barnes integrals~\cite{Freitas:2012sy};
second, the calculation of the vector and axial-vector couplings,
$g_A^f$ and $g_V^f$, now entirely relies on accurate
parametrisations~\cite{Hagiwara:1994pw,Hagiwara:1998yc,Cho:1999km,Cho:2011rk}; the
correction factors from a comparison with the Fortran ZFITTER
package~\cite{Arbuzov:2005ma,Bardin:1999yd} applied previously at very
high Higgs masses~\cite{Baak:2011ze} are no longer used.

\section{Results}

\label{sec:results}

The fit to all input data from Table~\ref{tab:results} converges with a global 
minimum value for the test statistics of $\ChiMin= \ChiMinVal$, obtained for
\NDFVal\ degrees of freedom. Using a pseudo Monte Carlo (MC) simulation and 
the statistical method described in~\cite{Flacher:2008zq} we find the 
$\ChiMin$ distribution shown in Fig.~\ref{fig:toys}. The resulting $p$-value
for the SM to describe the data amounts to \ProbVal (corresponding to 
$1.8\sigma$). This result is consistent with 
the naive $p$-value  ${\rm Prob}(\ChiMinVal,\NDFVal)= 0.08$. 

The inferior compatibility of the fit compared to earlier results~\cite{Baak:2011ze} 
is not primarily caused by the insertion of the new $M_H$ measurements in the fit, but 
is due to the usage of a more accurate $R^0_b$ calculation~\cite{Freitas:2012sy} 
that leads to a smaller SM prediction.\footnote
{
   The quantity of $R^0_b$ has only little dependence on $M_H$~\cite{Freitas:2012sy}.
}

\begin{table}[thbp] 
\newcommand{\bm}{\boldmath} \centering \setlength{\tabcolsep}{0.0pc}
{\small
\begin{tabular*}{\textwidth}{@{\extracolsep{\fill}}lccccc} 
\hline\noalign{\smallskip}
& & Free & Fit result & Fit result & Fit result incl. $M_H$  \\[-0.1cm]
\rs{Parameter} & \rs{Input value} & in fit & incl. $M_H$  & not incl. $M_H$  & but not exp. input in row \\
\noalign{\smallskip}\hline\noalign{\smallskip}
$M_{H}$ {\ft [GeV]}$^{(\circ)}$ & $125.7 \pm 0.4 $ & yes & $125.7 \pm 0.4$ & $94^{\,+25}_{\,-22}$ & $94^{\,+25}_{\,-22}$\\
\noalign{\smallskip}\hline\noalign{\smallskip}
$M_{W}$ {\ft [GeV]} &  $80.385\pm0.015$ & -- &  $80.367 \pm 0.007$ &  $80.380 \pm 0.012$ &  $80.359\pm0.011$\\
$\Gamma_{W}$ {\ft [GeV]} &  $2.085\pm0.042$ & -- &  $2.091\pm0.001$ &  $2.092\pm0.001$ &  $2.091\pm0.001$\\
\noalign{\smallskip}\hline\noalign{\smallskip}
$M_{Z}$ {\ft [GeV]} &  $91.1875\pm0.0021$ & yes &  $91.1878\pm0.0021$ &  $91.1874\pm0.0021$ &  $91.1983\pm0.0116$\\
$\Gamma_{Z}$ {\ft [GeV]} &  $2.4952\pm0.0023$ & -- &  $2.4954\pm0.0014$ &  $2.4958\pm0.0015$ &  $2.4951\pm0.0017$\\
$\sigma_{\rm had}^{0}$ {\ft [nb]} &  $41.540\pm0.037$ & -- &  $41.479\pm0.014$ &  $41.478\pm0.014$ &  $41.470\pm0.015$\\
$R^{0}_{\l}$ &  $20.767\pm0.025$ & -- &  $20.740\pm0.017$ &  $20.743\pm0.018$ &  $20.716\pm0.026$\\
$A_{\rm FB}^{0,\l}$ &  $0.0171\pm0.0010$ & -- &  $0.01627 \pm 0.0002$ &  $0.01637\pm0.0002$ &  $0.01624\pm0.0002$\\
$A_\ell$ $^{(\star)}$  & $0.1499\pm0.0018$ & --  & $0.1473^{\,+0.0006}_{\,-0.0008}$ & $0.1477 \pm 0.0009$ & $0.1468\pm 0.0005$$^{(\dagger)}$ \\
$\sinleff(Q_{\rm FB})$ &  $0.2324\pm0.0012$ & -- &  $0.23148^{\,+0.00011}_{\,-0.00007}$ &  $0.23143^{\,+0.00010}_{\,-0.00012}$ &  $0.23150\pm0.00009$\\
$A_{c}$ &  $0.670\pm0.027$ & -- &  $0.6680^{\,+0.00025}_{\,-0.00038}$ &  $0.6682^{\,+0.00042}_{\,-0.00035}$ &  $0.6680\pm0.00031$\\
$A_{b}$ &  $0.923\pm0.020$ & -- &  $0.93464^{\,+0.00004}_{\,-0.00007}$ &  $0.93468 \pm 0.00008$ &  $0.93463\pm0.00006$\\
$A_{\rm FB}^{0,c}$ &  $0.0707\pm0.0035$ & -- &  $0.0739^{\,+0.0003}_{\,-0.0005}$ &  $0.0740 \pm 0.0005$ &  $0.0738\pm0.0004$\\
$A_{\rm FB}^{0,b}$ &  $0.0992\pm0.0016$ & -- &  $0.1032^{\,+0.0004}_{\,-0.0006}$ &  $0.1036 \pm 0.0007$ &  $0.1034\pm0.0004$\\
$R^{0}_{c}$ &  $0.1721\pm0.0030$ & -- &  $0.17223\pm0.00006$ &  $0.17223\pm0.00006$ &  $0.17223\pm0.00006$\\
$R^{0}_{b}$ &  $0.21629\pm0.00066$ & -- &  $0.21474\pm0.00003$ &  $0.21475\pm0.00003$ &  $0.21473\pm0.00003$\\
\noalign{\smallskip}\hline\noalign{\smallskip}
$\mc$ {\ft [GeV]} &  $1.27^{\,+0.07}_{\,-0.11}$ & yes &  $1.27^{\,+0.07}_{\,-0.11}$ &  $1.27^{\,+0.07}_{\,-0.11}$ & -- \\
$\mb$ {\ft [GeV]} &  $4.20^{\,+0.17}_{\,-0.07}$ & yes &  $4.20^{\,+0.17}_{\,-0.07}$ &  $4.20^{\,+0.17}_{\,-0.07}$ & -- \\
$m_{t}$ {\ft [GeV]} &  $173.18\pm0.94$ & yes &  $173.52\pm0.88$ &  $173.14\pm0.93$ &  $175.8^{\,+2.7}_{\,-2.4}$\\
$\dalphaHadMZ$ $^{(\bigtriangleup\bigtriangledown)}$ &  $2757\pm  10$ & yes & $2755\pm  11$ & $2757\pm  11$ & $2716^{\,+  49}_{\,-  43}$\\
$\as(M_{Z}^{2})$ & -- & yes &  $0.1191\pm0.0028$ &  $0.1192 \pm 0.0028$ &  $0.1191\pm0.0028$\\
\noalign{\smallskip}\hline\noalign{\smallskip}
$\deltatheo M_W$ {\ft [MeV]}  & $[-4,4]_{\rm theo}$ & yes  & $4$ & $4$ & -- \\
$\deltatheo \sinleff$ $^{(\bigtriangleup)}$  & $[-4.7,4.7]_{\rm theo}$ & yes  & $-1.4$ & $4.7$ & -- \\
\noalign{\smallskip}\hline
\noalign{\smallskip}
\end{tabular*}
{\ft
$^{(\circ)}$Average of ATLAS ($M_H=126.0\pm0.4~(\rm{stat})\pm0.4~(\rm{sys})$) and CMS ($M_H=125.3\pm0.4~(\rm{stat})\pm0.5~(\rm{sys})$) measurements assuming no correlation of the systematic uncertainties (see discussion in Sect.~\ref{sec:input}).
$^{(\star)}$Average of LEP ($A_\ell=0.1465\pm0.0033$) and SLD ($A_\ell=0.1513\pm0.0021$) measurements, used as two measurements in the fit.
$^{(\dagger)}$The fit w/o the LEP (SLD) measurement gives $A_\ell=$ $0.1474^{\,+0.0005}_{\,-0.0009}$ ($A_\ell=$ $0.1467^{\,+0.0006}_{\,-0.0004}$).

$^{(\bigtriangleup)}$In units of $10^{-5}$.
$^{(\bigtriangledown)}$Rescaled due to $\as$ dependency.
}}

\captionof{table}{Input values and fit results for the observables and
  parameters of the global electroweak fit. The first and second
  columns list respectively the observables/parameters used in the
  fit, and their experimental values or phenomenological estimates
  (see text for references). The subscript ``theo'' labels theoretical
  error ranges.  The third column indicates whether a parameter is
  floating in the fit. The fourth column quotes the results of the
  complete fit including all experimental data. The fifth column gives
  the fit results for each parameter without using the $M_H$
  measurement in the fit. In the last column the fit results are given
  without using the corresponding experimental or phenomenological
  estimate in the given row.\label{tab:results} }
\end{table} 
\begin{figure}[t] 
\centering 
\includegraphics[width=11cm]{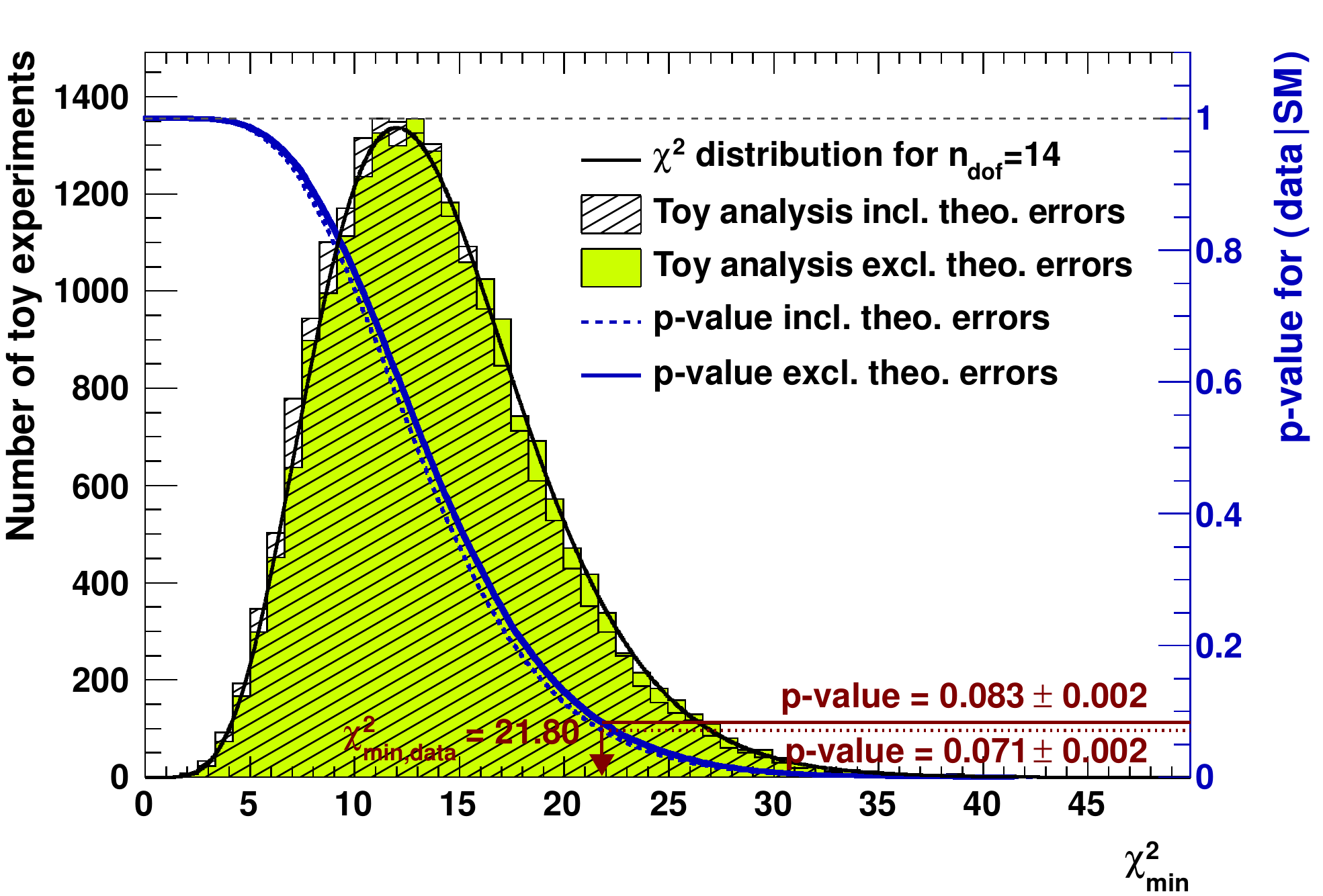} 
\caption[]{Result of a pseudo MC analysis of the complete electroweak fit. 
           Shown are distributions of the \ChiMin test statistics obtained from 
           a pseudo MC simulation with varying (hatched histogram) and fixed 
           theoretical uncertainty parameters ($\deltatheo$) 
           in the fit (shaded/green histogram). 
           The \ChiMin obtained in the complete fit to data
           is indicted by the arrow together with the $p$-values found
           for these
           two cases. Also shown is an idealised $\chi^2$ function assuming a 
           Gaussian case with 14 degrees of freedom (solid black line). }
\label{fig:toys}
\end{figure} 

\begin{figure}[t] 
\centering 
\includegraphics[width=7.0cm]{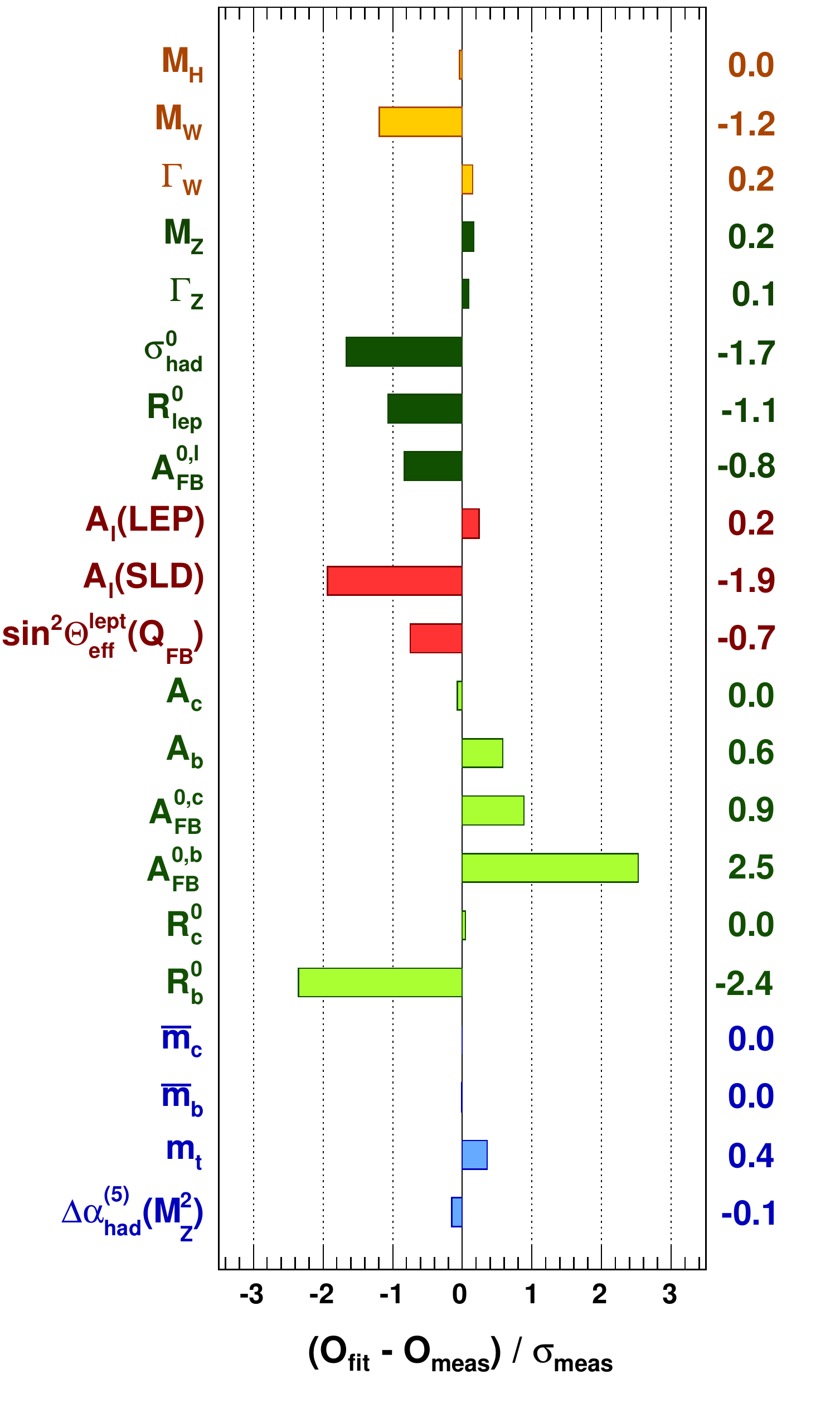} 
\includegraphics[width=9.2cm]{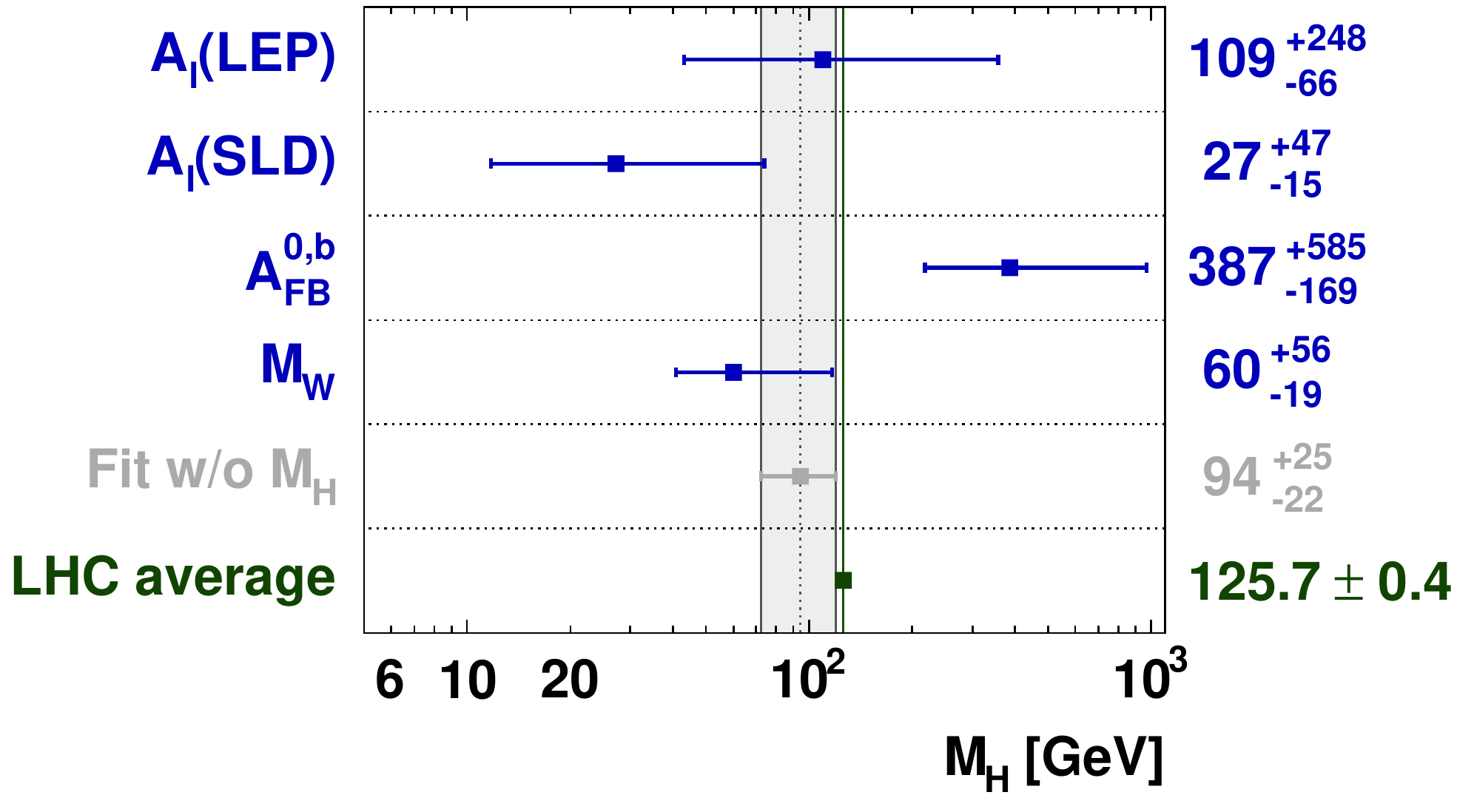}
\caption[]{Left: pull comparison of the fit results with the direct measurements in
           units of the experimental uncertainty. Right: determination of
           $M_H$ excluding the direct $M_H$ measurements and all the sensitive
           observables from the fit, except the one given. Note that the fit results 
           shown are not independent. }
\label{fig:pullplot}
\end{figure} 

The results of the complete fit for each fit parameter and observable 
are given in the fourth column of Table~\ref{tab:results}, together 
with their uncertainties estimated from their $\Delta\chi^2=1$ profiles. 
Figure~\ref{fig:pullplot} (left) shows the pull values obtained from the 
difference between the result of the fit and the input data in units of the 
data uncertainty. No single pull value exceeds $3\sigma$. The known 
tension between the left-right asymmetry and $A_{\rm FB}^{0,b}$ is reproduced. 
The new $R_b^0$ calculation~\cite{Freitas:2012sy} increases the discrepancy 
between the $R_b^0$ prediction and its measurement from $-0.8\sigma$ to $-2.4\sigma$.

The fifth column in Table~\ref{tab:results} gives the results obtained
without using the $M_H$ measurements in the fit (\ie, $M_H$ is a freely varying 
parameter). In this case, which represents the well known result of the 
standard electroweak fit prior to the $M_H$ measurement, the fit converges with a 
global minimum of $\ChiMin= 20.3 $ for 13 degrees of freedom 
(${\rm Prob}(20.3, 13) = 0.09$). We obtain
\begin{equation}
   M_H = 94^{\,+25}_{\,-22}\:\gev\;, 
\label{eq:mh}
\end{equation}
consistent within 1.3$\sigma$ with the $M_H$ measurements. The top left panel
of Fig.~\ref{fig:scans} displays the corresponding $\Delta\chi^2$
profile versus $M_H$ (grey band) compared to the new
$M_H$ measurements of ATLAS and CMS (red/orange data points) and the
$\Delta\chi^2$ profile of the fit including the $M_H$ measurement
(blue curve). 

Figure~\ref{fig:pullplot} (right) shows the determination of $M_H$
in fits in which among the four observables providing the strongest
constraint on $M_H$, namely $A_l$(LEP), $A_l$(SLD), $A_{\rm FB}^0$,
and $M_W$, only the one indicated in a given row of the plot is
included in the fit. For comparison also the indirect fit result
(without the $M_H$ measurement) and the direct measurement are shown
as vertical bands.

\begin{figure}[!t]
\epsfig{file=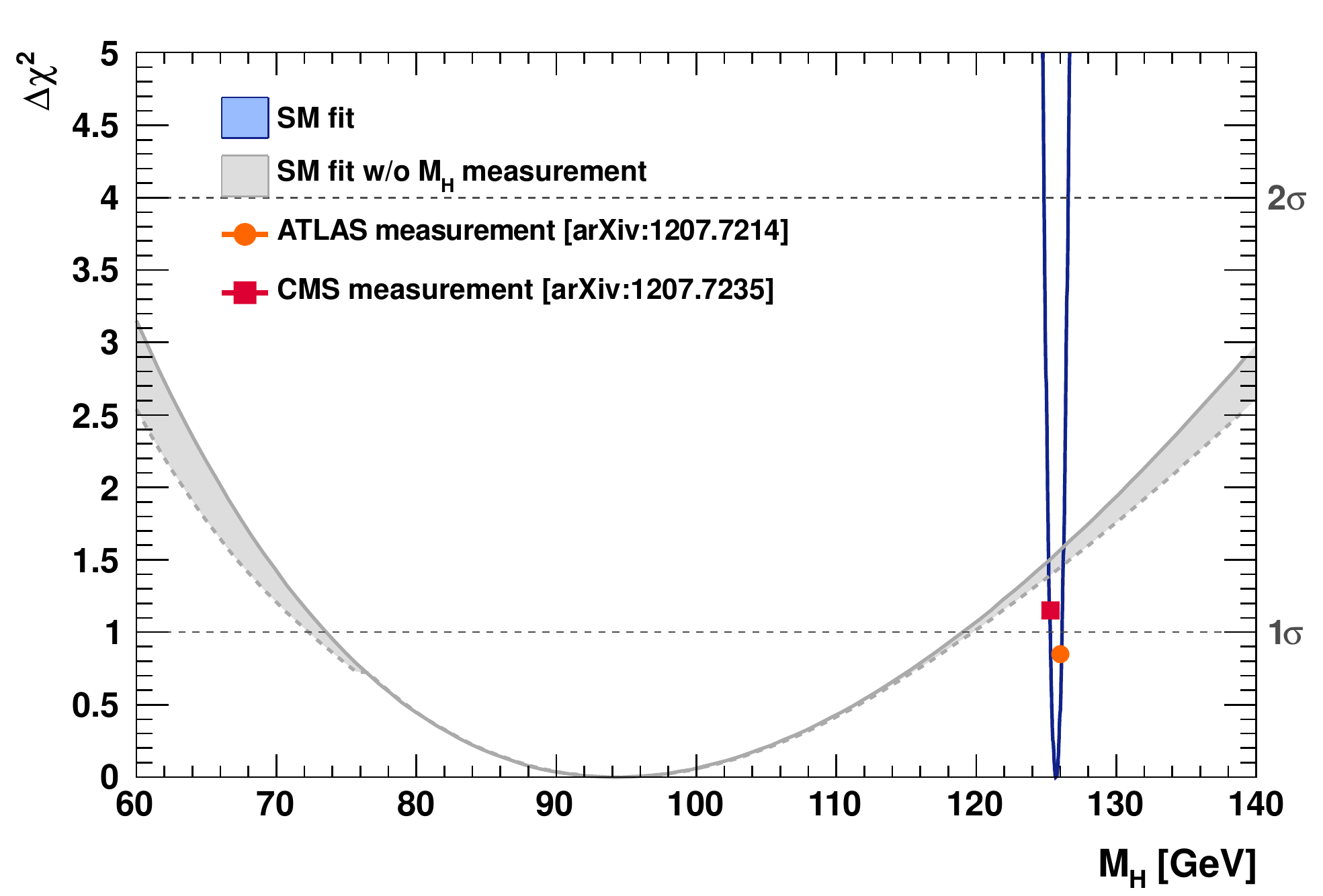, scale=0.38}
\epsfig{file=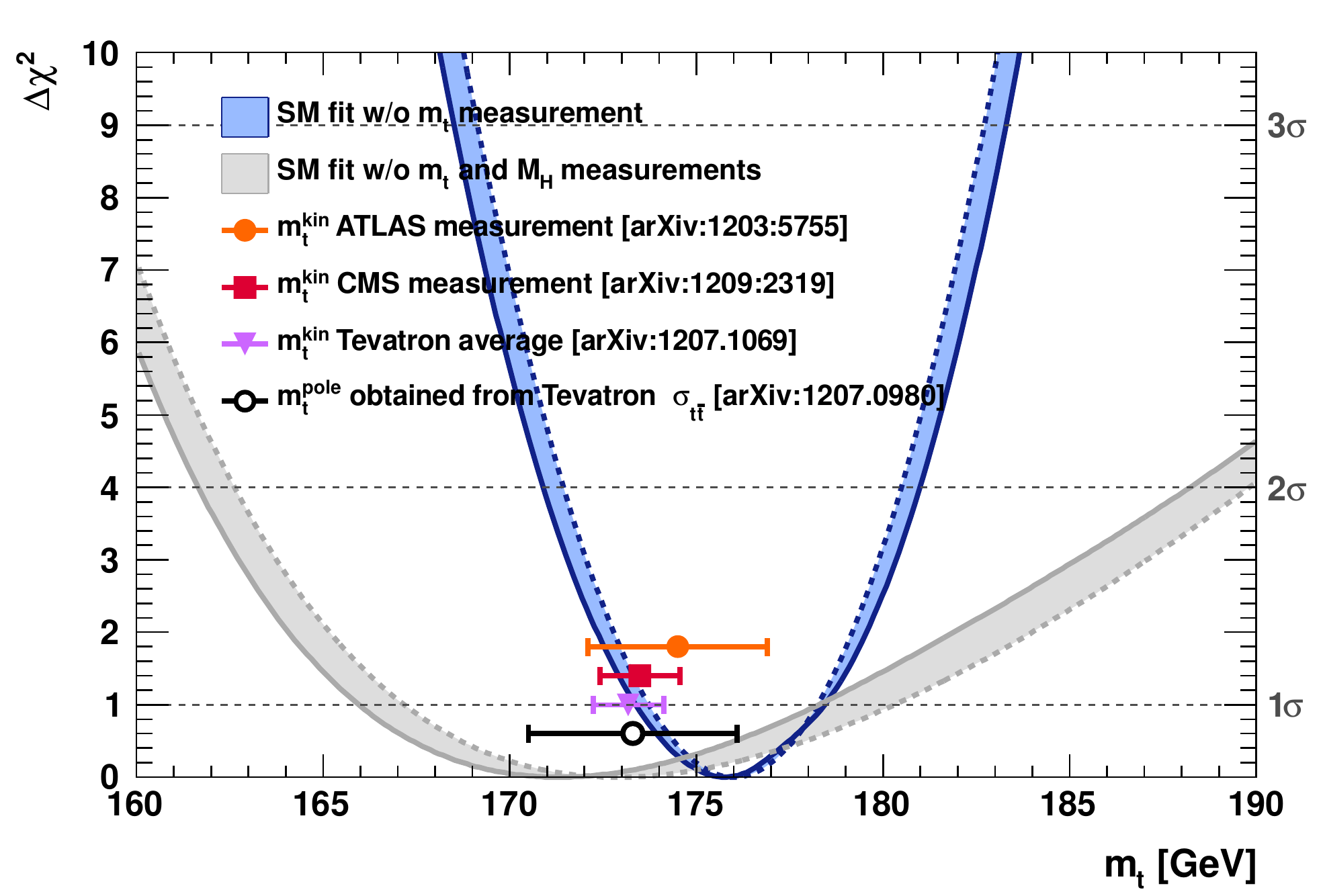, scale=0.38}
\epsfig{file=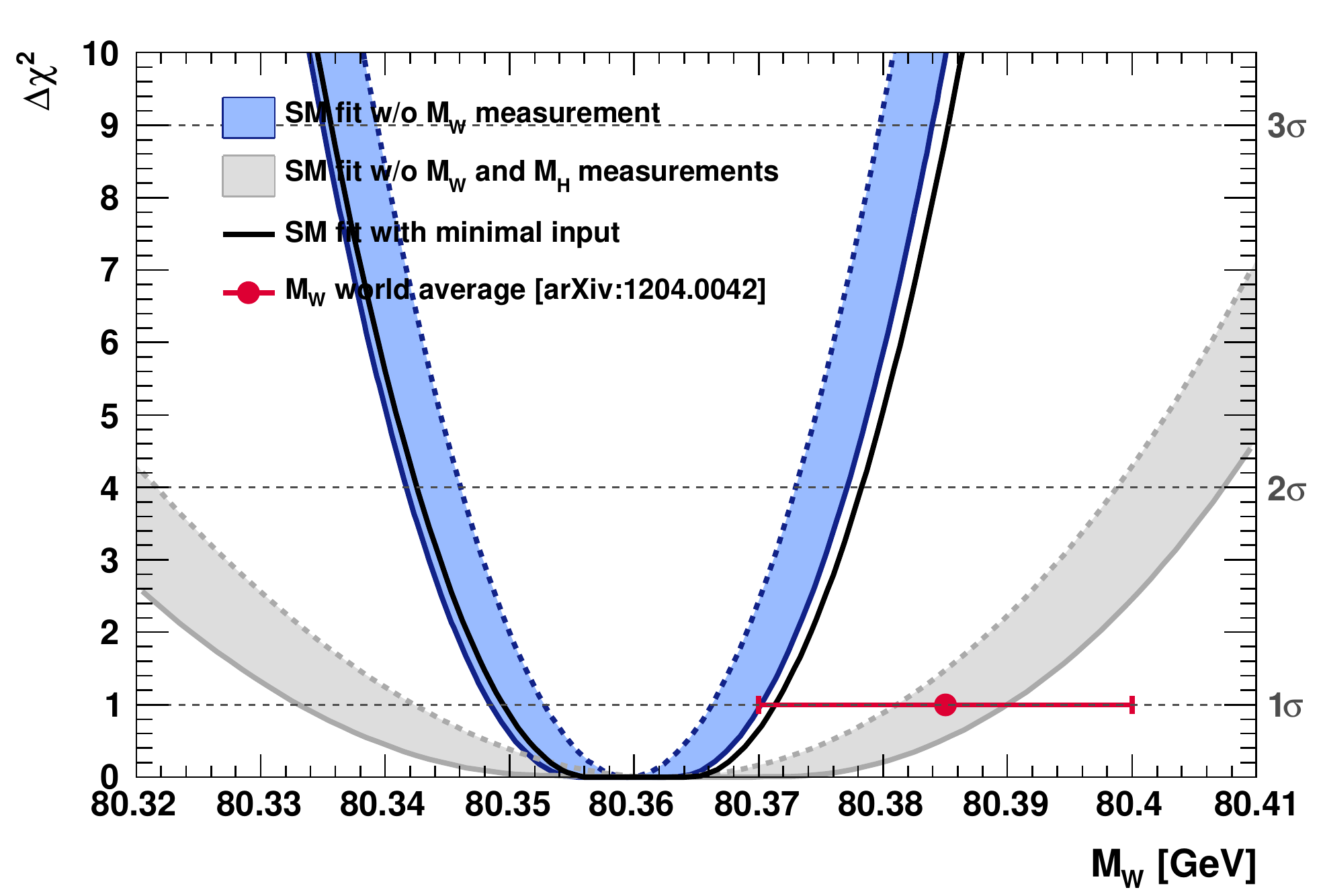, scale=0.38}
\hspace{0.95cm}
\epsfig{file=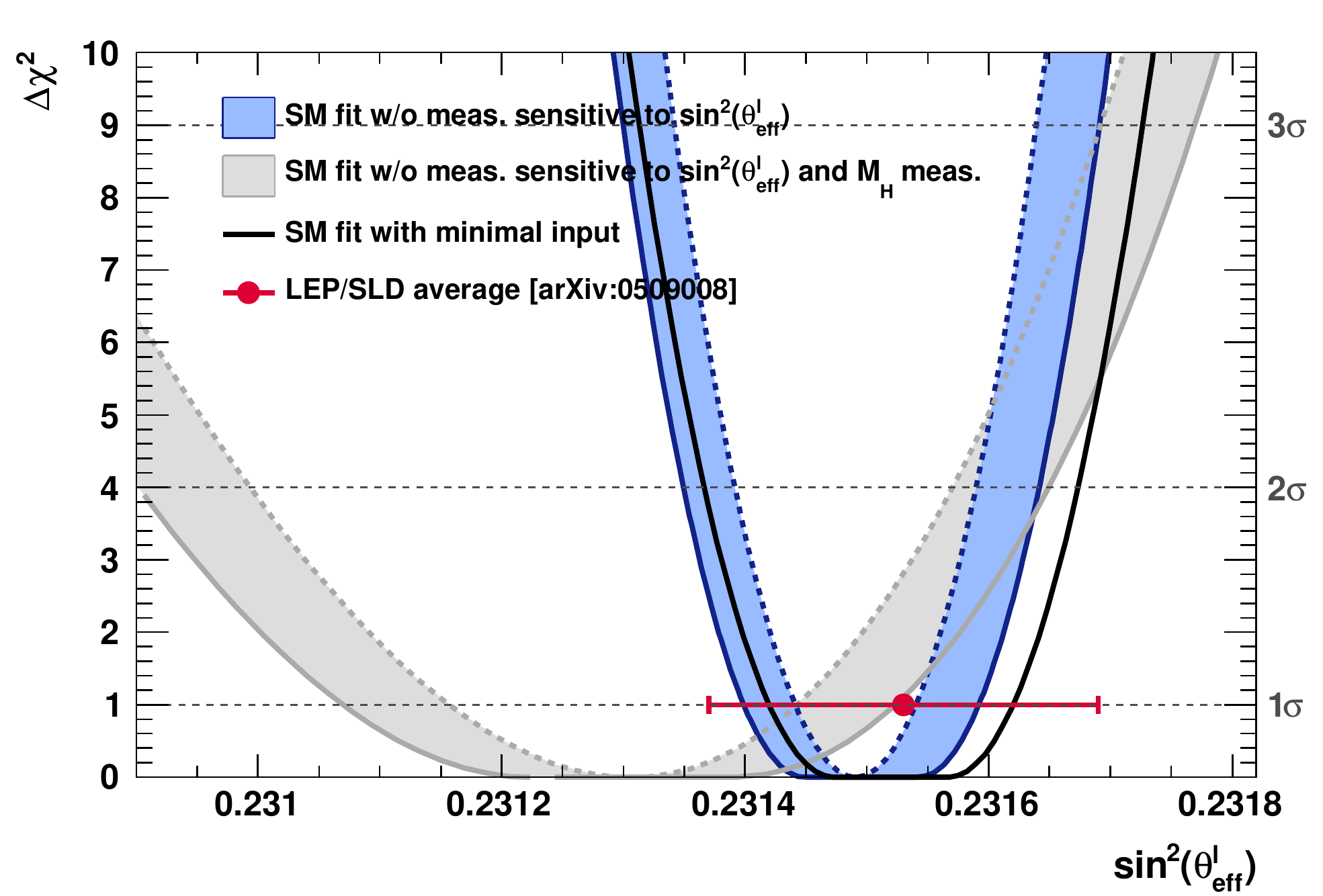, scale=0.38}
  \caption[]{$\DeltaChi$ profiles as a function of the Higgs mass (top
    left), the top quark mass (top right), the $W$ boson mass (bottom
    left) and the effective weak mixing angle (bottom right).  The
    data points placed along $\DeltaChi=1$ represent
    direct measurements of the respective observable and their $\pm
    1\sigma$ uncertainties.  The grey (blue) bands show the results
    when excluding (including) the new $M_H$ measurements from (in)
    the fits.  For the blue bands as a function of $m_t$, $M_W$ and
    $\sinleff$ the direct measurements of the observable have been
    excluded from the fit in addition (indirect determination). The
    solid black curves in the lower plots represent the SM prediction
    for $\sinleff $ and $M_W$ derived from the minimal set of input
    measurements, as described in the text.  In all figures the solid
    (dotted) lines illustrate the fit results including (ignoring)
    theoretical uncertainties in the fit.}
\label{fig:scans}
\end{figure}

The remaining plots in Fig.~\ref{fig:scans} show the $\DeltaChi$ profile
curves versus $m_t$ (top right), $M_W$ (bottom left), and $\sinleff$
(bottom right) obtained without using the corresponding experimental
measurement in the fit (indirect determination, \cf last column of
Table~\ref{tab:results}).\footnote 
{ 
  For the indirect determination of
  $\sinleff$, shown as the blue band in Fig.~\ref{fig:scans} (bottom
  right), we exclude from the fit all experimental measurements with
  direct sensitivity to $\sinleff$, namely the measurements of
  $\Gamma_{Z}$, $\sigma_{\rm had}^{0}$, $R^{0}_{\l}$, $A_{\rm
    FB}^{0,\l}$, $A_\ell$, $\sinleff(Q_{\rm FB})$, $A_{c}$, $A_{b}$,
  $A_{\rm FB}^{0,c}$, $A_{\rm FB}^{0,b}$, $R^{0}_{c}$ and
  $R^{0}_{b}$. As a compensation of the missing value of $R^{0}_{\l}$
  we provide a value for $\asZ$. Since the fit
  results are independent of the exact $\as$ value, we use our
  fit result $0.1191\pm 0.0028$ in this case.  
} 
For comparison also
the corresponding profile curves excluding in addition the new $M_H$
measurements are shown (grey bands).  The results from the direct
measurements for each variable are also indicated by
data points at $\DeltaChi=1$.\footnote
{
 We show the aforementioned result of the Tevatron combination of the
 direct top mass measurements~\cite{Aaltonen:2012ra}, the top pole
 mass derived from the measured $t\bar{t}$ cross-section at the
 Tevatron ($m_t=173.3 \pm 2.8$\,GeV), assuming no new physics
 contributes to this cross section measurement~\cite{Alekhin:2012sx},
 the direct top mass measurement of ATLAS determined in 1.04\,${\rm
   fb}^{-1}$ of $pp$ collisions at $\sqrt{s}=7$\,TeV ($m_t=174.5 \pm
 2.4$\,GeV)~\cite{ATLAS:2012aj}, the direct top mass measurement of
 CMS based on 5.0\,${\rm fb}^{-1}$ of $7$\,TeV data
 ($m_t=173.5 \pm 1.1$\,GeV)~\cite{CMSmt}, the aforementioned $W$ mass
 world average~\cite{TevatronElectroweakWorkingGroup:2012gb} and the
 LEP/SLD average of the effective weak mixing angle ($\sinleff=0.23153
 \pm 0.00016$)~\cite{:2005ema}.
\label{ftn:mtop}
}  
The insertion of $M_H$ substantially improves the precision 
of the fit predictions.

The fit indirectly determines the $W$ mass (\cf Fig.~\ref{fig:scans} -- bottom left, 
blue band) to be
\beqn
  M_W &=& 80.3593 
          \pm 0.0056_{m_t} \pm 0.0026_{M_Z} \pm 0.0018_{\Delta\alpha_{\rm had}} \\
      & & \phantom{80.3593} 
          \pm  0.0017_{\as} \pm 0.0002_{M_H} \pm 0.0040_{\rm theo}\,, \\[0.2cm]
      &=& 80.359 \pm 0.011_{\rm tot} \;,
\label{eq:mw}
\eeqn
which exceeds the experimental world average in precision. The different 
uncertainty contributions originate from the uncertainties in the input 
values of the fit as given in the second column in Table~\ref{tab:results}. 
The dominant uncertainty is due to the top quark mass. Due to the weak, logarithmic 
dependence on $M_H$ the contribution from the uncertainty on the Higgs mass 
is very small compared to the other sources of uncertainty. Note that in the 
\Rfit scheme~\cite{Hocker:2001xe,Charles:2004jd} the treatment of the 
theoretical uncertainty as uniform likelihood corresponds a linear addition
of theoretical and experimental uncertainties. Quadratic addition would 
give a total uncertainty in the $M_W$ prediction of 0.008.

The indirect determination of the effective weak mixing angle (\cf Fig.~\ref{fig:scans} 
-- bottom right, blue band) gives
\beqn
  \sinleff &=& 0.231496 
                 \pm 0.000030_{m_t} \pm 0.000015_{M_Z} \pm 0.000035_{\Delta\alpha_{\rm had}} \\
           & & \phantom{0.231496}
                 \pm 0.000010_{\as} \pm 0.000002_{M_H} \pm 0.000047_{\rm theo} \,, \\[0.2cm]
      &=& 0.23150 \pm 0.00010_{\rm tot} \;,
\label{eq:sin2t}
\eeqn
which is compatible and more precise than the average of the LEP/SLD 
measurements~\cite{:2005ema}. The total uncertainty is dominated by that from
$\Delta\alpha_{\rm had}$ and $m_t$, while the contribution from the uncertainty 
in $M_H$ is again very small. Adding quadratically theoretical and experimantal
uncertainties would lead to a total uncertainty in the $\sinleff$ prediction of $0.00007$.

Finally, the top quark mass, \cf Fig.~\ref{fig:scans} 
(top right, blue band), is indirectly determined to be
\begin{equation}
  m_t = \TopMassInd \;,
\label{eq:mt} 
\end{equation}
in agreement with the direct measurement and cross-section based determination
(\cf Footnote~\ref{ftn:mtop}). 

The measured value of $M_H$ together with the fermion masses, the
strong coupling strength $\asZ$ and the three parameters defining the
electroweak sector and its radiative corrections (chosen here to be
$M_Z$, $G_F$ and $\dalphaHadMZ$) form a minimal set of parameters allowing 
one, for the first time, to predict all the other SM
parameters/observables. A fit using only this minimal set of input
measurements\footnote 
{ 
  For $\asZ$ we use the result from Table~\ref{tab:results}.  
}  
yields the SM predictions $M_W=80.360\pm 0.011\:{\rm GeV}$ and 
$\sinleff = 0.23152 \pm 0.00010$. The
$\DeltaChi$ profile curves of these predictions are shown by the solid
black lines in Fig.~\ref{fig:scans} (bottom left) and (bottom right).
The agreement in central value and precision of these results with
those from Eq. (\ref{eq:mw}) and ({\ref{eq:sin2t}) (cf. blue bands in
  the plots) illustrates the marginal additional information provided
  by the other observables.

\begin{figure}[!t]
\begin{center}
\epsfig{file=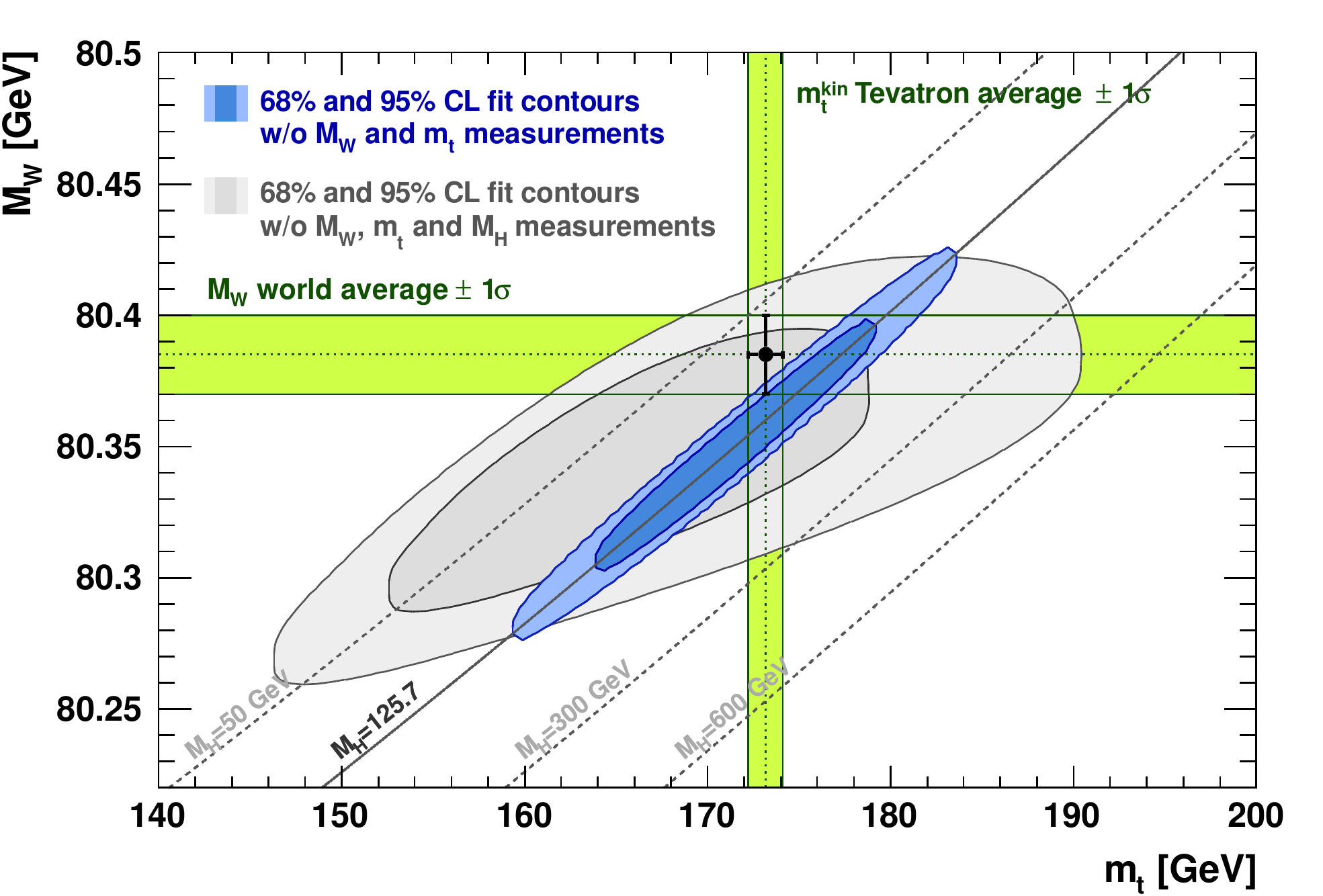, scale=0.6}
\end{center}
  \caption[]{ Contours of 68\% and 95\% CL obtained from scans
    of fixed $M_W$ and $m_t$. The blue (grey) areas illustrate the fit
    results when including (excluding) the new $M_H$ measurements. The
    direct measurements of $M_W$ and $m_t$ are always excluded in the
    fit. The vertical and horizontal bands (green) indicate the
    1$\sigma$ regions of the direct measurements.}
\label{fig:wvst}
\end{figure}

Figure~\ref{fig:wvst} displays CL contours of scans
with fixed values of $M_W$ and $m_t$, where the direct
measurements of $M_W$ and $m_t$ were excluded from the fit. The
contours show agreement between the direct measurements (green
bands and data point), the fit results using all data except the $M_W$, 
$m_t$ and $M_H$ measurements (grey contour areas), and the fit results 
using all data except the experimental $M_W$ and $m_t$ measurements (blue
contour areas). The observed agreement again demonstrates the impressive
consistency of the SM.

\begin{figure}[!t]
\begin{center}
\epsfig{file=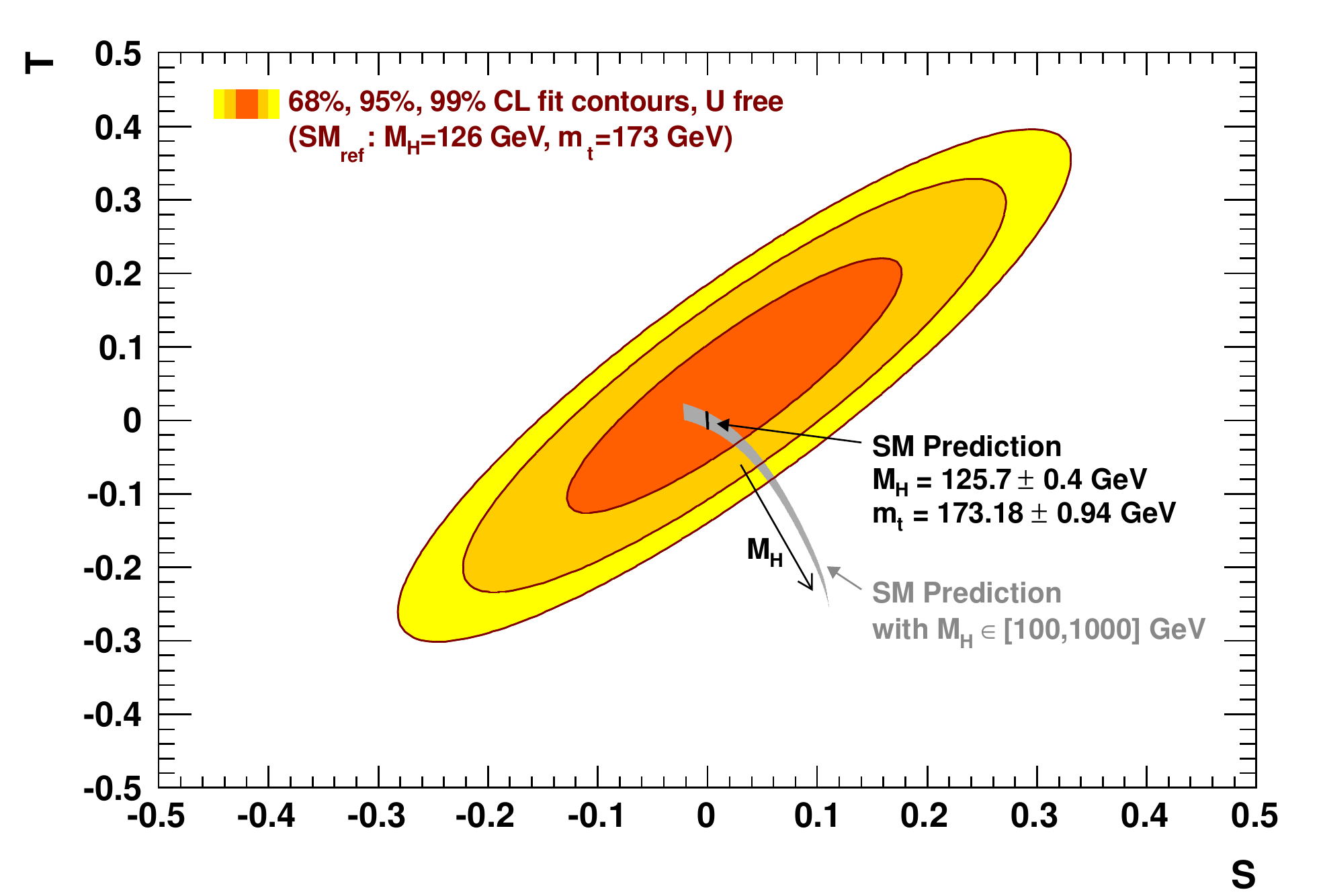, scale=0.4}
\epsfig{file=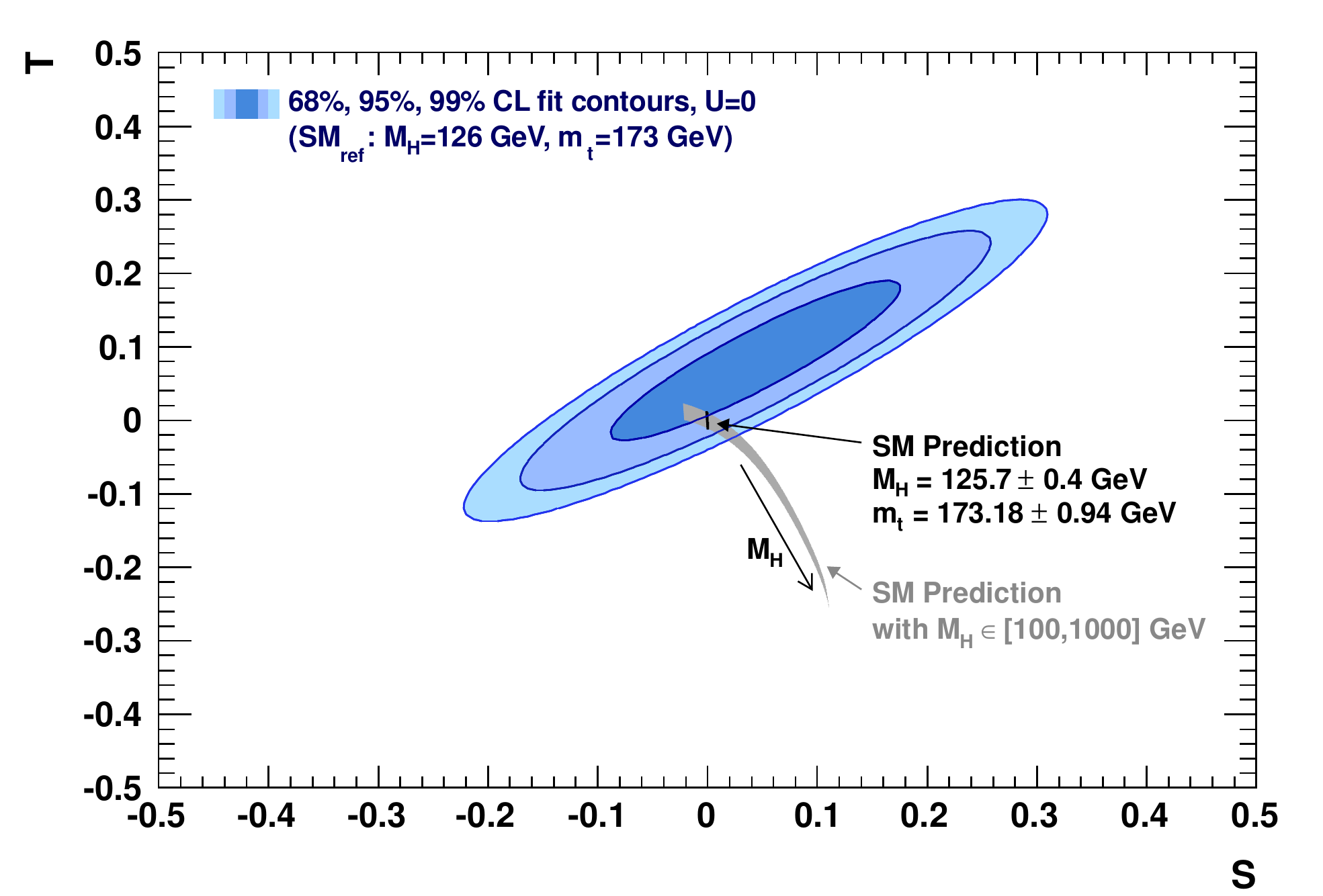, scale=0.4}
\end{center}
  \caption[]{ Experimental
    constraints on the $S$ and $T$ parameters with respect to the SM
    reference ($M_{H,\rm ref}=126$\:GeV and $m_{t,\rm
      ref}=173$\:GeV). Shown are the 68\%, 95\% and 99\% CL allowed
    regions, where the third parameter $U$ is left unconstrained
    (orange, left) or fixed to 0 (blue, right). The prediction in the SM is given
    by the black (grey) area when including (excluding) the new $M_H$
    measurements.  }
\label{fig:STU}
\end{figure}

Following the approach in \cite{Baak:2011ze} we extract from the
electroweak fit the \STU parameters~\cite{Peskin:1990zt,Peskin:1991sw}
describing the difference between the oblique vacuum corrections as
determined from the experimental data and the corrections expected in
a reference SM (SM$_{\rm ref}$ defined by fixing $m_t$ and $M_H$).
After the recent discovery, we change our definition of the reference
SM for the \STU calculation to $M_{H,\rm ref}=126$\,GeV and $m_{t,\rm
  ref}=173$\,GeV.  With these we find: 
\beq
  S= \SParam\:, \hspace{0.5cm}
  T= \TParam\:, \hspace{0.5cm}
  U=\UParam\:,
\eeq 
with correlation coefficients of $\STParamCor $
between $S$ and $T$, and $\SUParamCor$ ($\TUParamCor $) between $S$
and $U$ ($T$ and $U$). Fixing $U=0$ we obtain $S|_{U=0}= \SParamNU$
and $T|_{U=0}= \TParamNU$ with a correlation coefficient of
$\STParamCorNo$. Figure~\ref{fig:STU} shows
the 68\%, 95\% and 99\% CL allowed regions in the
$(S,T)$ plane for freely varying $U$ (left) and the 
constraints found when fixing $U=0$ (right). For illustration
also the SM prediction is shown. The $M_H$ measurement 
reduces the allowed SM area from the grey sickle, defined by letting
$M_H$ float within the indicated range, to the narrow black strip. 

\section{Conclusion}

Assuming the newly discovered particle at $\sim$126\:\gev 
to be the Standard Model (SM) Higgs boson, all fundamental parameters of the SM are 
known. It allows, for the first time, to overconstrain the SM at the electroweak 
scale and to evaluate its validity. The global fit to all the electroweak 
precision data and the measured Higgs mass results in a goodness-of-fit $p$-value
of 0.07. Only a fraction of the contribution to the ``incompatibility''
stems from the Higgs mass, which agrees at the 1.3$\sigma$ level with the fit prediction.
The largest deviation between the best fit result and the data is introduced by 
the known tension between $A_{\rm FB}^{0,b}$ from LEP and $A_\ell$ from SLD,
predicting respectively a larger (by 2.5$\sigma$) and smaller (1.9$\sigma$)
Higgs mass, and by $R^0_b$ for which an improved calculation increased the
deviation from the measurement from previously 0.8$\sigma$ to 2.4$\sigma$.

The knowledge of the Higgs mass dramatically improves the SM predictions of several
key observables. The uncertainties in the predictions of the $W$ mass, $\sinleff$, 
and the top mass decrease from 28 to 11\:\mev, $2.3\cdot10^{-5}$ to $1.0\cdot10^{-5}$,
and from 6.2 to 2.5\:\gev, respectively. The improved accuracy sets a 
benchmark for direct measurements, which has been reached (and surpassed) only 
for the top mass. Theoretical uncertainties due to unknown higher order electroweak 
corrections contribute approximately half of the uncertainties in the $M_W$ 
and $\sinleff$ predictions. 

The results reported in this letter depend on the validity of the assumption
that the observed particle is indeed the SM Higgs boson. New physics may lead
to deviations in the couplings, which also affect the global fit. The next round 
of experimental updates by ATLAS and CMS will lead to a more precise assessment
of the new particle's properties and are expected with great excitement. 

{\small\em
We thank Louis Fayard whose questions have triggered the more detailed error 
analysis for the $M_W$ and $\sinleff$ predictions provided in this version of 
the paper. 
}

\addcontentsline{toc}{section}{References}
\bibliography{References}{}

\end{document}